# Sonification of Facial Actions for Musical Expression

**Mathias Funk, Kazuhiro Kuwabara, and Michael J. Lyons**
ATR Intelligent Robotics & Communication Labs
2-2-2 Hikari-dai, Keihanna Science City
Kyoto 619-0288 Japan
mlyons@atr.jp

## ABSTRACT
The central role of the face in social interaction and non-verbal communication suggest we explore facial action as a means of musical expression. This paper presents the design, implementation, and preliminary studies of a novel system utilizing face detection and optic flow algorithms to associate facial movements with sound synthesis in a topographically specific fashion. We report on our experience with various gesture-to-sound mappings and applications, and describe our preliminary experiments at musical performance using the system.

## 1. INTRODUCTION
The idea of using facial expressions and intentional facial actions for the purpose of musical expression was conceived by one of us a few years ago in relation to an enjoyable, but unusual and therefore semi-private, form of play we call *face dancing,* in which the face is moved expressively in time with a music. While face dancing may not be a widely recognized art form in itself, stylized, controlled facial expression does play an important role in a variety of performing arts including mime and classical Indian dance. The appeal of using the face as a means of expression is related to the central role the face plays in verbal, non-verbal, and affective aspects of human social interaction including visual aspects of speech and non-verbal affective vocalizations, as well as facial expressions and gestures.

Our previous work [9,11,12] has been primarily focused on using the mouth for the purpose of musical control. In the present study we return to the general concept of a facial gesture interface and explore the expressive possibilities of various actions of the entire visible area of the face for interactive computer music. Using a larger area of the face leads us to relax requirements for detailed, precise control and aim rather for a transparent acoustic accompaniment (or sonification) of facial movement. We call the system a **So**nifier of **F**acial **A**ctions, abbreviated **SoFA**. As will be discussed, augmenting face-to-face interaction with sound suggests several experimental musical activities or performance scenarios which leverage upon human face-to-face interaction. Practically speaking, such a system may also help people to become consciously aware of their habits of facial action or it could augment visual processing of the facial actions of others. It might also find application as a support tool for facial physiotherapy exercises or for people with face processing disorders.

This paper is organized as follows. We briefly review closely related work in section 2. In section 3 we describe the prototype of a system for sonifying facial actions, in real-time video streams, in a topography specific fashion by combining optic flow estimation with automatic face detection. Facial movements generating optic flow having a supra-threshold magnitude trigger MIDI notes which are sent to a sound synthesis module. Section 4 describes our experiments using facial action to trigger various sound samples and outlines strategies for composition or musical activities which we are beginning to work up for performance purposes. The final section concludes the paper.

## 2. RELATED WORK
There has been relatively little research on the use of facial actions for intentional human machine interaction, with a few notable exceptions. Salem and Zhai [19] presented *Tonguepoint,* a Tongue pointing device in which a small pressure sensitive isometric joystick is mounted on a mouthpiece held with the teeth allowing cursor control using the tongue. Orio [17] devised a system to monitor the shape of the oral cavity by analyzing the response to acoustic stimulation which permitted independent control of two parameters. Vogt et al. [23] developed *Tongue 'n' Groove* which uses ultrasound imaging to measure the tongue position and movement in real time for real-time sound synthesis control.

Computer vision-based methods suggest the possibility of a non-contact, non-invasive method for acquiring human gestures from a distance. This area has grown recently, because of the increasing feasibility of robust, real-time vision systems for non-specialized hardware [18]. An earlier paper (NIME-03) reviewed the application of vision-based gesture analysis to the field of musical interfaces. To this we would like to add mention of the early, impressive work of Finnish electronic musician and computer pioneer Erkki Kurenniemi [8], whose DIMI-O (**DI**gital **M**usical **I**nstrument − **O**ptical) was probably the first digital video-based gestural musical interface. Written documentation of the DIMI-O and other DIMI musical interfaces is scant, however Kurenniemi's work was recently the subject of a documentary film which is recommended to anyone interested in human machine interfaces. The DIMI-O allowed digitized input from a video camera to modify the digital memory of a note sequencer attached to an electronic organ. This allowed a dancer to "play" note sequences remotely with her movements. Kurenniemi's group experimented with various forms of visual input, including movements of the face.

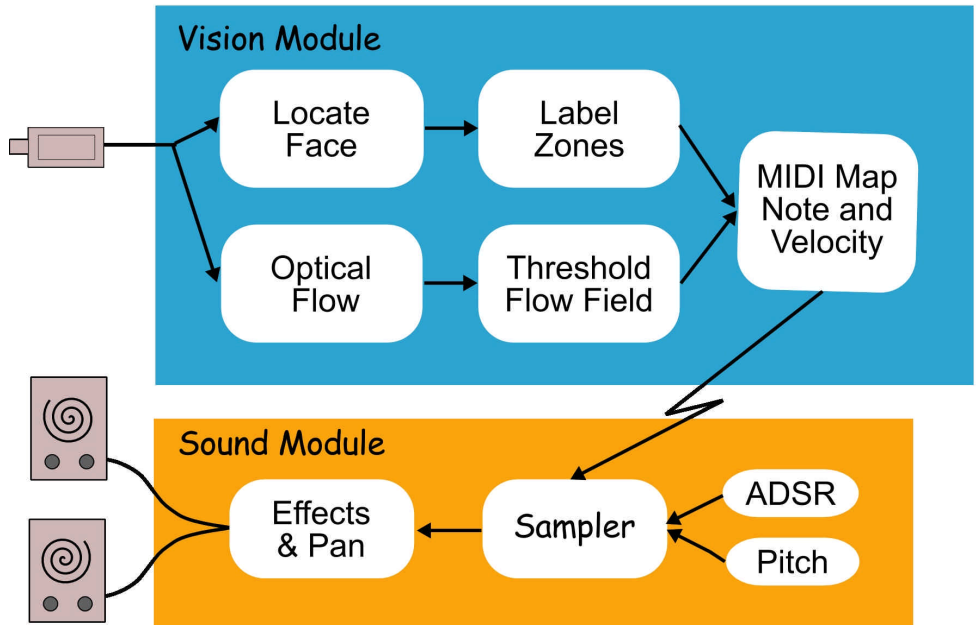

**Figure 1 Vision and Sound modules of the SoFA musical interface.**

In recent years several groups which have attempted to use video-based methods to acquire facial actions for musical expression. Lyons [9,11,12] developed the *Mouthesizer*, a vision-based system for extracting shape parameters of the external appearance of the mouth cavity such as height, width, and aspect ratio and mapping these to musical control parameters. Ng's *MvM* (Music via Motion) system used a simple deformable model of facial feature geometry to map facial movements to MIDI controls [15,16]. Merrill [14] used an infrared vision-based system to recognize head nods, tilts, and shakes and use these for discrete and continuous control of an effects unit. Hornof [7] demonstrated a system that uses infrared vision-based eye-tracking for musical performance.

## 3. DESIGN & IMPLEMENTATION

A system schematic is shown in Figure 1. Vision-based facial action acquisition and sound synthesis modules of SoFA are described separately in sub-sections below.

### 3.1 Optic Flow Calculation

We explore the use of dense optic flow as a feature for extracting information about actions of the face. Optic flow measures displacements of image areas due to motions of the camera or rigid or non-rigid motions of elements of the visual scene. For a fixed camera and constant, or slowly changing, illumination the optic flow field gauges the local motion of the visual scene. The basic theory of optic flow was established by Horn and others [6]. The extensive literature on methods for calculating optic flow has been reviewed in [1] and [2]. Dense optic flow was one of the earliest computer vision techniques applied to the automatic processing of facial movements [13] and it has been used in some studies of gesture-based human-computer interaction, including one study of a conducting interface [4]. However optic flow has not previously been investigated as a possible technique for sonifying facial actions.

The method we use for calculating optic flow is known as *block matching* or *block correlation*. This algorithm has been found to be quite stable to image noise compared to derivative estimation techniques [1,2]. With the block correlation technique, a region from one image frame is matched to a region of the same size in the subsequent frame. Matching is determined by minimizing a distance measure, in this case an $L^1$ measure, or the sum of the absolute values of differences between pixels in the matched regions. The velocity vector is calculated from the displacement of the block between frames.

With SoFA, optic flow is calculated on a grid of blocks of size 8 x 8 pixels covering the entire 320 x 240 pixel image. Hence, for each video frame, the calculation produces a 40 x 30 array of two dimensional velocity vectors. SoFA calculates these optic flow fields at a frame rate of 15 fps assuming a standard Pentium processor running Windows XP and image acquisition via a USB 2 camera. At this frame rate feature displacements due to facial actions are up to a few pixels per time step.

The vector field is spatially smoothed by averaging over nearest neighbours to reduce effects of noise. Facial movements produce vector fields with greater spatial correlation than those produced by pixel noise. Noise therefore tends to be reduced by the spatial averaging procedure. Empirical verification with a selection of typical facial actions showed that the resulting vector fields corresponded well with observed motion fields of facial parts. Figure 2 illustrates the optic flow field calculated for a sample facial action.

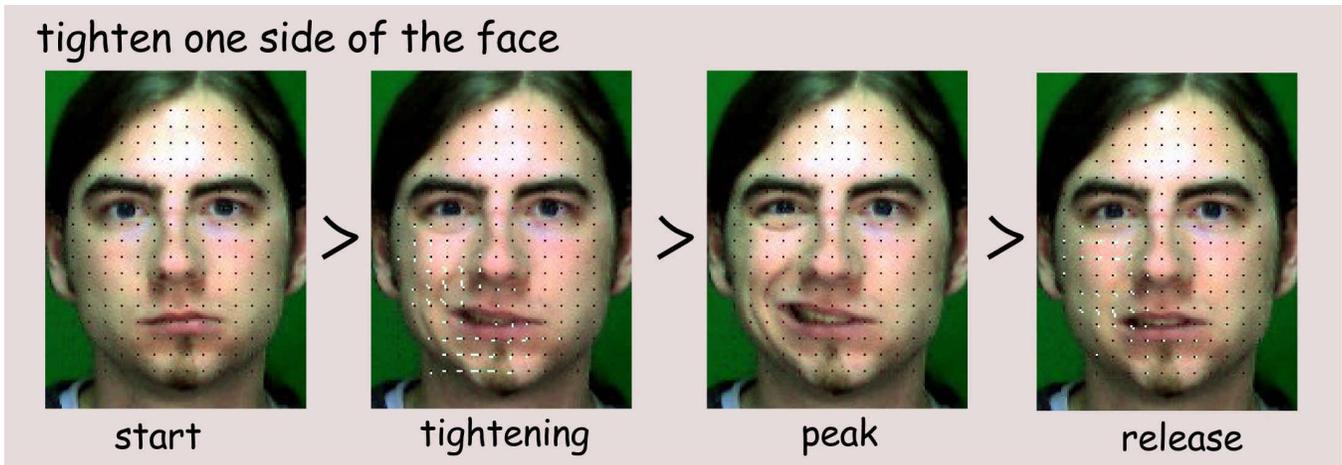

**Figure 2 Sample facial action with associated optic flow vector fields illustrated as white line segments.**

### 3.2 Face Detection Algorithm

The optic flow calculation yields a grid of motion vectors which covers the entire image. Our goal is a system which responds to facial movements in a topography-specific fashion. Therefore we need to determine whether a motion vector corresponds to a region of the image containing a face. Furthermore we would like to associate motion vectors with parts of the face.

Considerable research effort has been devoted to methods for detecting faces in single images and image sequences (see [5] for a survey). Of the many approaches available, we choose a powerful and widely popular algorithm due to Viola and Jones [22], known as cascade-correlation. Details of the algorithm are given in the original paper. The algorithm automatically registers a rectangle of fixed aspect ratio with the head with the face centered in the interior. The Viola-Jones face detection algorithm is robust to changes in scale of the face as well as in the level of illumination. It detects faces accurately for out of plane rotations of up to roughly 30 degrees and roughly the same amount of in plane rotation of the head.

Seven approximate facial zones are calculated from the position of the facial region of interest (ROI) and are used to label the motion vectors. The seven regions, shown in figure 3, are: left and right parts of the brow, immediately above the eyebrow, the left and right eyes; left and right cheeks; and the mouth. Each facial zone has a salient motion detection rule which evaluates the x and y components of the optic flow vector and determine whether or not to trigger a motion event. For example, the eyebrow regions only evaluate the vertical component of the optic flow vectors. The 8 salient motion vectors having the highest magnitude in a given frame generate MIDI note-on events. A MIDI note-off is sent after a fixed number (n=4) of video frames has been processed. No notes are triggered unless the size of the detected face is greater than a certain minimum value.

### 3.3 Visualization

Each motion vector of supra-threshold magnitude in one of the facial zones is represented in the rendered video output of the face as a white line segment indicating the optic flow at that point. Grid points which result in a MIDI output are indicated with a small coloured circle which expands until the note-off signal is sent. This visual representation is intended to resemble the expanding ripples that result when a pebble is thrown into a pond. The four zone types are colour coded: the eyebrows region is visualized with red circles, the eye region with yellow circles, cheeks with blue, and mouth with green circles.

### 3.4 MIDI Events

The pitch of MIDI notes generated depends on the vertical coordinate of the triggering grid point, while the velocity is determined by the magnitude of the optic flow vector. A pentatonic scale is used, with the lowest note (MIDI note 40), at the bottom of the video frame and the highest note at the top of the frame (MIDI note 112). Notes from each facial zone are sent on a separate MIDI channel, resulting in use of a total of 7 channels.

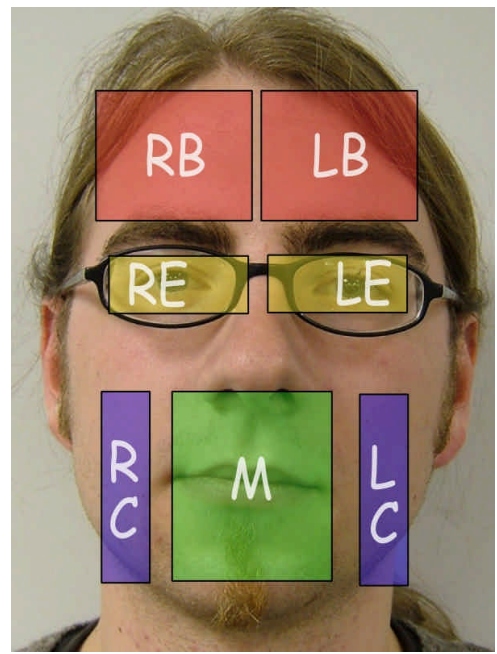

**Figure 3**. **Facial Zones used to trigger MIDI events.**

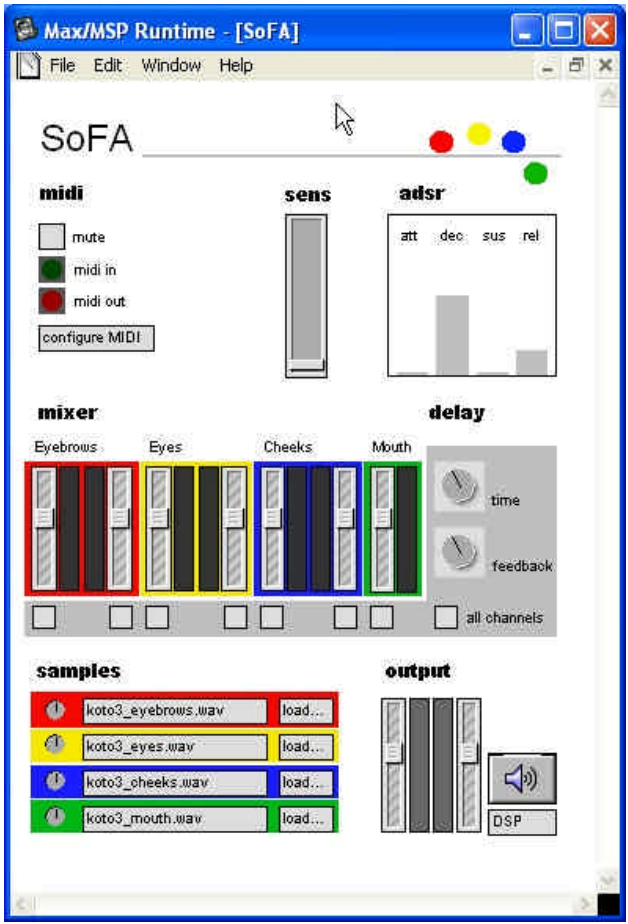

**Figure 4 SoFA synthesis module controls.**

## 3.5 Synthesis Module

For musical synthesis we use a bank of seven polyphonic software samplers implemented in a Max/MSP patch and executed with the runtime environment on the Wintel machine which runs the vision module. Analogous parts of the right and left sides of the face share one of four available samples which are colour-coded red, yellow, blue and green. The samples are associated with the four types of facial zones: brow, eyes, cheeks, and mouth. Each channel has the value of stereo pan set to its relative topography position according to the topography of the face. This creates an auditory space which corresponds to the spatial map of motor actions on the face. Sound output is mixed down to a single stereo channel.

Each sample can be processed with a switchable independent delay effect unit. This is intended to enable compositions having both figure and ground. Sustained sound textures, or drones, may be started with some facial zones, while ephemeral patterns are triggered with other zones. Feedback in the delay unit is low pass filtered and the delay time ranges from zero to two seconds. A single set of parameters controls all enabled delay effects.

## 3.6 Synthesis Module Control Interface

A screen shot of the synthesis control interface may be seen in figure 4. We decided to restrict the number of adjustable controls in the synthesis module in order to encourage performers to concentrate on playing by working with their facial movements. Most of the sampler parameters are set as global values. For example, the same ADSR envelope is used for all four samples. It is hoped this will avoid the temptation to use the control interface to tweak individual sounds such as is common in laptop performances.

A default set of percussive samples is provided at startup and (wav format) samples may be loaded to play different compositions. Independent left and right audio channel volume controls are provided for each sample to allow matching of audio levels during setup.

MIDI input may be toggled with the space bar to turn facial input on and off without interrupting audio synthesis. A global audio mute control is also provided.

## 4. APPLICATIONS

We believe that with the appropriate refinement and development SoFA could find several therapeutic uses:

- For ambient sonification of one's facial movements as one engages in a solitary activity such as reading, working at a computer, meditating etc… This is a form of biofeedback can increase awareness of habitual behaviour patterns.

- As a tool for supporting facial physiotherapy. Cerebral aneurysms, Parkinson's and other afflictions often result in facial paralysis which is treated with facial exercises. An evolved version of SoFA could be used to provide feedback about exercise as well as make it more engaging.

- SoFA could be useful for those suffering prosopagnosia or other forms of brain-damage which compromise face or facial expression recognition capabilities and interfere with social interaction expertise.

- SoFA might allow the blind to "hear" facial expressions.

SoFA poses some very interesting possibilities for musical expression and we are currently exploring the use of the system for solo and collaborative performance.

In solo play, SoFA provides an acoustic counterpart to a performance of facial mime. Most interesting to us, however, are the possibilities of SoFA in collaborative music making [3]. Two, or possibly more, performers communicate non-verbally via facial mime as well as emotional expressions. SoFA adds a sonic layer to the interaction which can itself become a means of non-verbal communication with each other as well as with the audience. SoFA then becomes a scaffold for the construction of *artificial expressions,* of novel forms of non-verbal communication which are enabled by the fact of machine-mediation.

Two complementary aesthetic considerations have played a role in the design of SoFA and in our ongoing exploration of the musical possibilities of the system. They are *Empathy* and *Ostranemie.*

*Empathy*, or shared feeling, can evolve when there exists a physical basis for *mimesis* [10,21]. To engage mimetic interaction we have attempted to create a transparent action-to-synthesized sound mapping based on the anatomical topography of the face.

*Ostranemie,* or defamiliarization, is an aesthetic concept introduced by Russian formalist Viktor Shklovsky (1893-1984) [20]. The idea is that by representing everyday things in novel, unconventional ways one can make the perception of the familiar take on new aspects. Attention becomes focused on the *act of perception* itself, which leads to new insights into the commonplace. By augmenting facial expression with a sonic layer, we focus attention on these quasi-conscious acts and bring them to conscious awareness.

We are experimenting with several compositions in a genre that might be described as *expressive* **concrète** with a view to eventual public performance in a few months time.

## 5. CONCLUSION

This paper has discussed the design and implementation of a novel interface which allows a sonic layer to be associated with facial actions and expressions. The system, known as SoFA, is sufficiently stable and robust to be used in performance. We are currently exploring the musical possibilities of the SoFA and hope to use it for a public performance in a few months time.

## 6. ACKNOWLEDGMENTS

This work was supported in part by the National Institute of Information and Communications Technology.